%% file: main.tex
\documentclass[conference,a4paper,final]{IEEEtran}
\input{content/macro}

\usepackage{flushend}

\begin{document}

\author{%
\IEEEauthorblockN{Renlord Yang$^{\ast\mathsection}$, Toby Murray$^\ast$, Paul
Rimba$^\mathsection$, Udaya Parampalli$^\ast$}\\
\IEEEauthorblockA{
  $^\ast$School of Computing and Information Systems, University of Melbourne, Victoria, Australia\\
  $^\mathsection$Data61/CSIRO, Sydney, Australia\\
  $^\ast$rnyang[at]student.unimelb.edu.au, {toby.murray,udaya}[at]unimelb.edu.au\\
  $^\mathsection$first.last[at]data61.csiro.au\\
}
}

\thispagestyle{empty}

\title{Empirically Analyzing Ethereum's Gas Mechanism}

\maketitle
\input{content/abstract}

\input{content/introduction}
\input{content/background}
\input{content/methodology}
\input{content/result}

\input{content/discussion}
\input{content/solution}
\input{content/related}
\input{content/conclusion}

\bibliographystyle{IEEEtran}
\bibliography{content/references/references.bib}
\end{document}

%% file: content/macro.tex
\usepackage{amsmath,amssymb,amsfonts}
\usepackage{algorithmic}
\usepackage{epsfig,endnotes}
\usepackage{mathtools}
\usepackage{graphicx, fixltx2e, float, tikz, adjustbox, url, hyperref, booktabs}
\usepackage{pgfplots}
\usepackage{xcolor}
\usepackage{cleveref}
\usepackage{xspace}

\usepackage{ifthen}
\newcommand{\mycomment}[2]{NOTE:***#1***: #2}
\newboolean{showcomments}
\setboolean{showcomments}{true} 
\ifthenelse{\boolean{showcomments}}
{\renewcommand{\mycomment}[2]{
    \fbox{\bfseries\sffamily\scriptsize COMMENT #1}
        {\sf\small$\Rightarrow$\textit{#2}$\Leftarrow$}
    }
}
{\renewcommand{\mycomment}[2]{}}

\newcommand{\eg}{\latinlocution{e.g.,} \xspace}
\newcommand{\ie}{\latinlocution{i.e.,} \xspace}

\newcommand{\latinlocution}[1]{\textit{#1}}

\newcommand{\abs}[1]{\left|#1\right|}

\usepackage[group-separator={,}]{siunitx}

%% file: content/abstract.tex
\begin{abstract}
  Ethereum's Gas mechanism attempts to set transaction fees in accordance with
  the computational cost of transaction execution: a cost borne by default by
  every node on the network to ensure correct smart contract execution.
  Gas encourages users to author transactions that are efficient
  to execute and in so doing encourages node \emph{diversity}, allowing
  modestly resourced nodes to join and contribute to the security of the
  network.

  However, the effectiveness of this scheme relies on Gas costs being correctly
  aligned with observed computational costs in reality.  In this work, we
  performed the first large scale empirical study to understand to what degree
  this alignment exists in practice, by collecting and analyzing Tera-bytes
  worth of nanosecond-precision transaction execution traces.  Besides
  confirming potential denial-of-service vectors, our results also shed light
  on the role of I/O in transaction costs which remains poorly captured by the
  current Gas cost model. Finally, our results suggest that under the current
  Gas cost model, nodes with modest computational resources are disadvantaged
  compared to their better resourced peers, which we identify as an ongoing
  threat to node diversity and network decentralization.
\end{abstract}

%% file: content/introduction.tex
\section{Introduction}
\label{sec:introduction}
Cryptocurrency networks have seen massive growth and innovation, due to their
decentralised nature and extensibility. The Ethereum network is a prime
example, whose design aims to maximise extensibility while preserving
decentralisation. Ethereum adopts Bitcoin's proof-of-work distributed consensus
mechanism and supports the execution of complex transactions through
smart-contracts written in the Ethereum Virtual Machine (EVM) instruction set.

Given the gravity of decentralisation, \emph{node diversity}---the ability for
heterogeneous machines of varying capabilities to participate in the network---is
a key consideration in the design of the EVM and setup of consensus
parameters.
In particular, Ethereum encourages anyone to freely join the network and
participate even with modest computational resources without unduly hindering
the network's transaction throughput.  Achieving decentralisation while
maintaining diversity and throughput in an ecosystem of heterogeneous machines
with differing computing capabilities is non-trivial because, by default, all
network nodes are required to execute and check the results of each new
transaction, to derive the current state of the network.  Indeed, we argue that
having a diversity of nodes able to check correct transaction execution is
crucial for the network's security, especially against the possibility of
colluding miners.

To manage the cost of having nodes execute each new transaction,
when a user creates a new transaction in Ethereum, the user must pay a fee for
the network to execute the transaction.  Part of this fee represents an
approximation of the computational and storage cost of running the transaction.
This fee is paid in virtual units, called Gas.  The Gas mechanism therefore
incentivises nodes to create transactions that are quick to execute as users
optimise towards paying as little gas as possible.  In addition, a block
\emph{gas limit} bounds the amount of gas that can be consumed by the
transactions in any single block of Ethereum's blockchain, approximating a
per-block upper bound on the computational complexity of transaction execution
and verification.  In this way, Ethereum balances the freedom to create complex
transactions against the need to ensure that transactions can be efficiently
verified.

If gas is to maximally incentivise nodes to minimise variable computational
costs, it is critical that it accurately reflects the real costs of transaction
execution.  To this end, the Ethereum Foundation has defined precise formulae
for calculating gas costs for transactions, as a function of the transaction
bytecode instructions~\cite{wood2014ethereum}.  Specifically, the current costs
were defined by setting an ideal ratio~$R$ between execution time and units of
gas on a single unidentified benchmark
machine\footnote{\url{https://docs.google.com/spreadsheets/d/1n6mRqkBz3iWcOlRem\_mO09GtSKEKrAsfO7Frgx18pNU/edit\#gid=0}}.
By doing so each EVM instruction opcode was assigned a fixed gas cost that
approximated its computational cost on the benchmark machine, calibrated
relative to the other instructions.

The accuracy of this calibration directly influences the utility of Gas for
incentivising transaction efficiency, and ensuring Ethereum can continue to
maintain decentralization in the absence of trust in the face of increasing
smart contract complexity.  To evaluate the gas mechanism in its mission to
price transactions effectively with respect to transaction execution time, we
perform an in-depth empirical analysis of the Ethereum gas mechanics in an
attempt to determine how they operate under real world usage scenarios.  Our
study was carried out by measuring transaction execution times across an
interval of 2+ million blocks (post-EIP150) in the Ethereum blockchain, replaying all state
transitions that have occurred since the genesis block.  This allowed us to
compute the observed Time-to-Gas ratio for each opcode, and their distributions,
shedding light on the calibration precision of the Gas mechanics.  Our
measurements were collected and contrasted between two machines: one highly
efficient machine representing an upper bound on computational resources of
individual Ethereum nodes and the other a typical desktop machine. Our
contributions are as follows:
\begin{itemize}
\item A detailed collection of EVM traces for each transaction,
  with opcode execution times, measured on two disparate
  machines with defined specifications.
\item We identify that some EVM opcodes are mispriced, particularly those
  involving disk I/O, presenting potential
  denial-of-service vectors.
\item  We analyse the effects of I/O and the impact of
  in-memory caching strategies
  employed by the reference client to mitigate I/O costs.
\item We show that the current gas cost model favours more performant machines,
  which we identify as an ongoing threat to node diversity and decentralization.
\item We briefly discuss possible solutions to these issues.
\end{itemize}

%% file: content/background.tex
\section{Background}
\label{sec:background}
In this section, we will provide the building blocks for understanding how the
Gas Mechanism operates in Ethereum.

\subsection{Ethereum}
\label{sec:background-ethereum}
The accounting unit for usage within the Ethereum network is known as Ether.
Transactions in Ethereum utilise the \emph{account} model of computation where
each account is a reference to either a smart contract or a payment account.
Both payment and smart contract accounts may send and receive Ethers.

The \emph{Ethereum Virtual Machine} (EVM) is an interpreter which executes
transactions that invoke smart contracts on the blockchain.  Every participant
in the Ethereum network operates an instance of the EVM and executes each new
transaction.  Doing so ensures that no node can lie about the results of a
transaction undetected.  The EVM implements the consensus rules that govern the
validity of transactions, which includes the gas cost for each operator and
other details (\eg the block gas limit) that define valid state transitions.

The ordering of transactions is defined by their position in the blockchain and
they are executed serially.  Parallel verification of
transactions~\cite{yu2017smart} is non-trivial due to the possibility of
side-effects via state mutation.  As per other decentralised consensus network,
Ethereum uses \emph{Nakamoto Consensus} to order state transitions for its
network state.  Miners continuously solve proof-of-work puzzles to include valid
transactions into blocks.  All participants, miners and ordinary users alike
verify every new transaction to independently derive the state of the network.

\subsection{Transaction and State}
\label{sec:transaction}
Transactions describe state transitions in Ethereum.  There are three types of
transactions in Ethereum: transfer, message-call, and contract creation
transactions.  \emph{Transfer} transactions transfer Ether balances from one
account to another without the usage of the interpreter.  \emph{Message-call}
transactions contain data payload which is used an input to execute smart
contracts published on the Ethereum blockchain.  \emph{Contract creation}
transactions publish new smart contracts onto the Ethereum blockchain. Accepting
a transaction in Ethereum requires each participating node on the network to
first verify the cryptographic seals and replaying the transaction.  During the
execution runtime for message-call transactions, gas is consumed for each
executed EVM instruction.

Ethereum nodes across the network update their node state each time a new block
is received.
During the update process, nodes update various data structures stored locally
which are used to verify future transactions, such as the state trie, account
trie, transaction trie, and transaction receipts trie, which are all necessary
to derive future block headers.  The \emph{state trie} is an record of the
current state of all accounts and implemented as an authenticated data
structured.  Where possible, to mitigate I/O disk access, some elements of the
state trie may be stored in memory for faster storage and retrieval to verify
future transactions (see \Cref{sec:results}).

When joining the network, participating users have two ways to obtain the
current state of the network.  Firstly, they can bootstrap their nodes by
performing \emph{Initial Blocks Download} (IBD), aka a ``full-sync'', to
independently derive the current state of the network.  During this process, the
node must verify the cryptographic elements and replay all transactions for each
block.  Replaying transactions at the time of writing takes many weeks of serial
computational time. Secondly, users can instead elect to obtain state snapshots
from other peers (\eg \emph{fast-sync}~\cite{fast-sync} in \texttt{geth} and
\emph{warp-sync} in \texttt{parity}). Doing so allows the node to avoid having
to replay all prior transactions. However, this comes at the price of having to
trust in the honesty of a majority of its peers---a non-trivial trust assumption
in the face of Sybil attacks~\cite{marcus2018low}.

\subsection{Gas: Computation Virtual Unit}
\label{sec:background-gas}
\emph{Gas}~\cite{eth-whitepaper,wood2014ethereum} is a virtual unit that is consumed
when computation is performed by the EVM and exist to align computational cost
to payable transaction fees.
In effect, it discorages authored transactions which are computationally
expensive to execute.
When transactions are submitted to miners to be included in the blockchain, the
author pays upfront a fixed amount of Gas.
As part of the payment process, the author also bids for a gas price denominated
in Ethers.
The transaction fee for an authored transaction is a product of the amount of
gas and the bidded gas price.

For each computational step performed by the miner during verification, Gas is
subtracted from the paid Gas.  An Out-of-Gas exception is thrown by the EVM when
Gas runs out.  Such transactions are still included in the blockchain as proof
that execution was attempted, and the gas paid is forfeited by the miner.  On
the other hand, valid transactions that have leftover gas will be refunded to
transaction author.

\subsection{Attacks}
\label{sec:attacks}
\newcommand{\EXTCODESIZE}{\texttt{EXTCODESIZE}}
Attacks against decentralized cryptocurrency networks have been increasing over
the past years due to its infancy and increased usage.  On the networking layer,
these networks are susceptible to eclipse attacks, first explained
by~\cite{singh2006eclipse} prevents participants from consensus
participation~\cite{heilman2015eclipse, marcus2018low}.  On the EVM layer,
numerous attacks~\cite{atzei2017survey} of varying severity exist.  Notably,
denial-of-service attacks mounted in the form of a resource exhaustion attack.
Resource exhaustion attacks were first identified in~\cite{luu2015demystifying} exploiting the fact that arbitrary
inclusion of long-running transactions may be included by problem givers
imposing an asymmetric amount of effort required to verify and execute these
transaction to determine if the output is correct.  Its impact when exploited
with our observations in the EVM Gas Mechanism causes a significant reduction in
transaction throughput and degradation in user experience as transactions take
significantly longer to execute.  Previously, between block 2.3M and 2.46M, a
severe mispricing of gas cost for the \EXTCODESIZE{} opcode~\cite{eth-ddos2016}
resulted in significantly longer transaction execution times across numerous
blocks which resulted in the EIP150 hard-fork to introduce re-aligned gas cost
for a number of EVM opcodes.

%% file: content/methodology.tex
\section{Methodology}\label{sec:methodology}
As an attempt to evaluate the Gas Mechanism,we take time measurements on
a per-opcode basis while transactions are executed by the EVM on two disparate
machines. In this section, we describe the hardware, software, procedures and
workloads undertaken to collect the measurements.

\subsection{Hardware Specification}\label{sec:hw-spec}
The first machine (A) used for experiment is a Dell R740xd and its setup is a
dedicated server instance without any hypervisors nor guest VM instances.
\Cref{tab:machinespec} lists the hardware specifications. While second machine
(B),  Dell Optiplex machine, representing a typical desktop node.

\begin{table}[h!]
  \caption{Experimental Platforms}
  \label{tab:machinespec}
  \centering
  \begin{tabular}{l l l}
    \toprule
    Machine & A & B \\ \midrule
        Storage Type & PCIe NVMe SSD & SATA3 SSD \\ \midrule
        CPU & Intel\textregistered\ Xeon\textregistered & Intel\textregistered\ Core\texttrademark \\ & Platinum &
          i7-4770@3.40GHz \\ & 8180M@2.50GHz \\ \midrule
        Threads (Core) & 2 (112) & 1 (4) \\ \midrule
        Cores (Socket) & 28 (56) & 4 (4) \\ \midrule
        Sockets & 2 & 1 \\ \midrule
        Memory & 1.5TB DDR4 & 16GB DDR3 \\ & 2300MHz & 1600MHz\\ \midrule
        OS & Ubuntu 16.04 LTS & Ubuntu 16.04 LTS \\ \midrule
        Kernel & Linux 4.15.0-33 & Linux 4.15.0-33 \\
        \bottomrule
  \end{tabular}
\end{table}

These two different machines were chosen with the intent of allowing our results
to extrapolated across the spectrum on typical Ethereum nodes at the time of
writing. The key difference between our hardware choice and community-run
Ethereum nodes is that most nodes are deployed in Virtual Private Server
settings within data centres~\cite{Kim:2018:MEN:3278532.3278542}.

\subsection{Software}\label{sec:software}
The software implementation used is the C++ Ethereum reference implementation,
also known as \texttt{aleth}\footnote{\url{https://github.com/ethereum/aleth}}
or \texttt{cpp-ethereum} previously.  For compilation options, we compiled with
default settings which uses LevelDB\footnote{\url{http://leveldb.org/}} as the
overlay DB back-end to manage the blockchain data and the state trie.
We elected to use the \texttt{C++} implementation as it is one of the reference
implementations maintained by the Ethereum Foundation and it features relatively
less runtime overhead compared to other implementations (\ie \texttt{geth,
ethereumj}).  Our methodology could be readily adapted to alternate software
implementations, and doing so we leave as future work.

\subsection{Setup}\label{sec:exp-setup}
Our setup involve collecting nanosecond-precision transaction execution traces.
Each transaction execution trace records the sequence of EVM opcodes
(instructions) executed by the transaction and, for each element of the
sequence, the length of time consumed by its execution in nanoseconds.
Execution traces were output to a file through buffered I/O and then following
the data collection were loaded into a database for further analysis.  All data
processing was performed outside of transaction execution to ensure that it did
not directly interfere with our timing measurements.  The logging of timing
results was done using line buffered output stream.  Intermittent buffer flushes
occur outside of the runtime of EVM opcode instances guaranteeing the buffer
flushes did not distort our time measurements.  Our measurements exclude all
time spent on other work such as the SHA3 computation for cryptographic
transaction validation as we are only concerned with the performance of the EVM
while executing transactions. We also omitted measurements for balance transfer
transactions.

To collect our data, we forked the reference client\footnote{at commit:
develop/aa73807c9a6f79114d36ce738658dcba0bc7fbb3} to include the necessary code
to perform timing measurements, making use of the standard \texttt{chrono} C++
library and user static tracing (UST) probing tool, \texttt{lttng}.

The reference C++ client that we forked makes use of various in-memory
software caches to avoid disk I/O. To understand their impact on transaction
execution, we also added code to measure, for each transaction, the number of
cache hits and misses.  We logged cache hits and misses for in-memory caches
\begin{itemize}
  \item Code Size Cache --- caches code size of most recently used smart
    contract.
  \item Empty Account Cache --- caches recently looked-up empty accounts.
  \item Account Cache --- caches recently read account balances.
  \item Code Cache --- caches recently used contract codes.
\end{itemize}

After the transaction traces were collected, we computed the the
Time-to-Gas ratio for each opcode and performed statistical analysis for
each opcode to visualise variability of the distribution which is shown in
\Cref{tab:muppet-table} and \Cref{tab:loula-table}.

\subsubsection{Noise}\label{sec:noise}
Any software measurement experiment is often subjected to noise due to the
internal and competing processes in a system. We note that our experimental
measurements cover a continuous data collection period that spans multiple weeks
of real time.  Our measurements cover transaction executions across 2 million
blocks where the same set of EVM opcodes are repeatedly executed millions, if
not billions of times.  Therefore, we anticipate that measurement noise due to
system resource contention will present as outliers in the data sets.

\subsubsection{System under test (SUT)}\label{sec:sut}
Machine A is a dedicated server instance.  During the data collection period,
the Ethereum client was the only heavy workload process with metrics collected
in a file for each transaction. Writes to the files were performed using
buffered I/O with pre-allocated memory in the heap.  The dedicated server
instance is a multi-tenanted instance, therefore there are other processes that
may have contended for IO access.  However, we contend that the noise as a result
of contention will be smoothed out with the high amount of collected traces.
Machine B is a personal desktop machine.  During the data collection period, the
Ethereum client was run within a Linux control group with a RAM hard-cap limit
of 8GB.  Numerous other processes that contend with hardware resources were also
running at the same time such as a web-browser, productivity tools, etc. We
elected to use a noisy setup for Machine B as it is representative of the
hardware choice used by a consumer user.

\subsection{Workloads}\label{sec:workload}
We ran two types of workloads for the purpose of collecting EVM traces.  From
\Cref{sec:background}, we noted that Ethereum opcodes require access to state
which is stored on disk, therefore we also measured the time spent for low-level
function calls to the client database interfaces.

First, the initial block download workload is used to gather measurements on all
transactions since the genesis block up to block 4.7M.  Between block 2.3M and
2.46M, a severe mispricing of gas cost for EVM opcodes that requires Disk I/O
resulted in a denial of service attack.  Discussed earlier in
\Cref{sec:attacks}, EIP150~\cite{eip150} was implemented to permanently
increase the cost of I/O-intensive opcodes to better align the verification time
against gas cost. All transaction measurement data prior to EIP150 has been
excluded from the results.  This workload mimics a typical use scenario where
transactions are continuously executed.

Second, we used \texttt{lttng}~\cite{desnoyers2006lttng} to collect more
detailed traces of function calls in the database interface implemented by the
client in an interval between two arbitrary block height.  This technique
enabled us to analyse internal function calls to the database interface and the
time spent for each call to associated database interface functions.  In
contrast to other tracing tools, such as Systemtap and \texttt{uprobes} with
perf, we found that instrumenting the client with \texttt{lttng} traces enabled
timing measurements with more precision.

%% file: content/result.tex
\section{Results}\label{sec:misalignment}

In this section we present our measurements from the collected
transaction traces of time and gas used during transaction execution.

\subsection{Baseline}\label{sec:baseline}
As mentioned in \Cref{sec:introduction} the gas cost for each opcode was set in
the consensus rule based on baseline measurements undertaken by the Ethereum
foundation.  The Ethereum foundation performed benchmarks to measure the
approximate execution time of each opcode on an unidentified host machine. Then
they set the gas cost for each opcode so that one unit of gas would represent
approximately 1$\mu$s (1,000 nanoseconds) worth of computation, on that machine.
Under this calibration, the baseline Time-to-Gas (aka Time/Gas) ratio of
execution time in nanoseconds to gas used for each opcode is 1,000. Smaller
values of this ratio represent better gas economy.

\subsection{Observations}\label{sec:results}
For each opcode we calculate its observed Time-to-Gas ratio for each of its
executions in our recorded traces, by dividing its observed execution time
by the amount of gas it is defined to consume under the current Gas cost model,
mentioned above.

A total of 1TB worth of traces were collected on both SUTs and the ratios for
each individual execution of each opcode was collected, and then grouped
according to opcodes. Doing so allowed us to produce a full distribution of
observed Time-to-Gas ratios for each opcode.  \Cref{tab:muppet-table} and
\Cref{tab:loula-table} present a range of summary statistics that capture each
distribution\footnote{More comprehensive results in a companion website is
available at
\url{https://github.com/renlord/bookish-octo-barnacle/blob/master/SNB2019.md}}.

The mean, $\mu$ is calculated by taking the average of the summation of all
measured time-to-gas ratios.  $Q_{1}$ and $Q_{3}$ respectively refer to the 25th
and 75th percentile in data.  The inter-quartile range (IQR) is calculated by
taking the difference of $Q_{3} - Q_{1}$. The Mean Absolute Deviation (MAD), a
robust statistical method to compute variability in data~\cite{huber2011robust},
$\sigma$ was computed using the method, $(\sum\limits_{i=1}^n \abs{x_{i} -
Q_{2}}) \div n$.  The Coefficient of Variability
(CoV)~\cite{abdi2010coefficient}, $\sigma \div \mu$ is used to measure the
relative spread between different opcodes with respect to their individual
distribution of observed ratios.  Values in parentheses are computed by subtracting the
mean/median against the baseline ratio of 1,000ns/gas and dividing by the original value,
to measure how far the computed mean/median has deviated from the
baseline.  The number of observations for each opcode has also been included as
a sanity check confirming that both machines executed the same exact
transactions for our workload. \dag~~denotes opcodes that have high execution
overhead due to the need to access the state trie.  For the semantics of each
EVM opcode, we refer the reader to the Ethereum yellowpaper~\cite{wood2014ethereum}.

\begin{table*}[htbp]
    \caption{Summary statistics for 10 selected opcodes by the highest mean
    ratio on Machine A as described in \Cref{tab:machinespec}.}
    \centering
    \begin{tabular}{l | l l | l l l l l l l}
        \toprule
       & \multicolumn{7}{l}{Machine A} \\ \cmidrule(lr){2-9}
        Opcode &
        $\mu$ & $Q_{2}$ (Median) & $Q_{1}$ & $Q_{3}$ & IQR & \#Observations & MAD $\sigma$ &
        CoV $\sigma$ $\div$ $\mu$ \\
        \midrule
        BLOCKHASH$^\dag$ & \num{34343} (3,334\%) & 110 (-89\%) & 56.0 & \num{87027.0} & \num{86971} & \num{169537} & \num{34284.32} & 1.00 \\
        SLOAD & 921 (-7\%) & 32 (-96\%) & 1.0 & 148.0 & 147 & \num{138736099} & 917.96 & 1.00 \\
        BALANCE$^\dag$ & 867 (-13\%) & 596 (-40\%) & 509.0 & 692.0 & 183 & \num{8644400} & 400.32 & 0.46 \\
        ORIGIN & 300 (-69\%) & 289 (-71\%) & 269.0 & 313.0 & 44 & \num{78807} & 34.22 & 0.11 \\
        ADDRESS & 297 (-70\%) & 297 (-70\%) & 262.0 & 318.0 & 56 & \num{5123311} & 38.67 & 0.13 \\
        COINBASE & 286 (-71\%) & 271 (-72\%) & 250.0 & 312.0 & 62 & \num{58315} & 38.73 & 0.13 \\
        CALLER & 267 (-73\%) & 261 (-73\%) & 223.0 & 299.0 & 76 & \num{47434902} & 46.54 & 0.17 \\
        MSTORE & 144 (-85\%) & 134 (-86\%) & 117.0 & 162.0 & 45 & \num{364731080} & 31.53 & 0.22 \\
        CALLDATALOAD & 142 (-85\%) & 134 (-86\%) & 89.0 & 192.0 & 103 & \num{68615078} & 49.18 & 0.34 \\
        EXTCODESIZE$^\dag$ & 142 (-85\%) & 4 (-99\%) & 2.0 & 27.0 & 25 & \num{12386619} & 141.00 & 0.99 \\
        CODESIZE & 95 (-90\%) & 53 (-94\%) & 45.0 & 58.0 & 13 & \num{36583} & 54.69 & 0.57 \\
        SGT & 94 (-90\%) & 58 (-94\%) & 37.0 & 117.0 & 80 & \num{308277} & 56.52 & 0.60 \\
        MLOAD & 94 (-90\%) & 76 (-92\%) & 69.0 & 95.0 & 26 & \num{248684463} & 25.54 & 0.27 \\
        DIV & 94 (-90\%) & 89 (-91\%) & 42.0 & 135.0 & 93 & \num{84993868} & 47.39 & 0.50 \\
        SDIV & 85 (-91\%) & 64 (-93\%) & 38.0 & 106.0 & 68 & \num{162142} & 48.00 & 0.56 \\
        GASPRICE & 61 (-93\%) & 48 (-95\%) & 44.0 & 54.0 & 10 & \num{492471} & 19.71 & 0.32 \\
        PUSH5 & 56 (-94\%) & 58 (-94\%) & 41.0 & 65.0 & 24 & \num{355623} & 13.91 & 0.24 \\
        CALLDATASIZE & 50 (-94\%) & 50 (-95\%) & 42.0 & 55.0 & 13 & \num{26912010} & 10.12 & 0.20 \\
        MULMOD & 48 (-95\%) & 17 (-98\%) & 12.0 & 80.0 & 68 & 3,279 & 36.29 & 0.75 \\
        SLT & 46 (-95\%) & 42 (-95\%) & 37.0 & 49.0 & 12 & \num{4399360} & 10.36 & 0.22 \\
        \bottomrule
    \end{tabular}
    \label{tab:muppet-table}
\end{table*}

\begin{table*}[htbp]
    \caption{Summary statistics for 10 selected opcodes by the highest mean
    ratio on Machine B as described in \Cref{tab:machinespec}.}
    \centering
    \begin{tabular}{l | l l | l l l l l l l}
        \toprule
       & \multicolumn{7}{l}{Machine B} \\ \cmidrule(lr){2-9}
        Opcode &
        $\mu$ & $Q_{2}$ (Median) & $Q_{1}$ & $Q_{3}$ & IQR & \#Observations & MAD $\sigma$ &
        CoV $\sigma$ $\div$ $\mu$ \\
        \midrule
        BLOCKHASH$^\dag$ & \num{35156} (3,415\%) & 117 (-88\%) & 54.0 & \num{81575.0} & \num{81521} & \num{169537} & \num{35097.84} & 1.00 \\
        SDIV & \num{26854} (2,585\%) & 85 (-91\%) & 48.0 & 126.0 & 78 & \num{162142} & \num{26805.20} & 1.00 \\
        SLOAD & \num{16808} (1,580\%) & 31 (-96\%) & 0.0 & 201.0 & 201 & \num{138736099} & \num{16805.73} & 1.00 \\
        BALANCE$^\dag$ & \num{12883} (1,188\%) & 7,070 (607\%) & 5,592.0 & 8,823.0 & 3,231 & \num{8644400} & 7,808.86 & 0.61 \\
        SGT & \num{12499} (1,149\%) & 61 (-93\%) & 45.0 & 136.0 & 91 & \num{308277} & \num{12454.36} & 1.00 \\
        EXTCODESIZE$^\dag$ & 2,369 (136\%) & 4 (-99\%) & 2.0 & 26.0 & 24 & \num{12386619} & 2,367.62 & 1.00 \\
        SLT & 1,459 (45\%) & 85 (-91\%) & 53.0 & 173.0 & 120 & \num{4399360} & 1,404.95 & 0.96 \\
        DIV & 663 (-33\%) & 120 (-88\%) & 43.0 & 175.0 & 132 & \num{84993868} & 613.12 & 0.92 \\
        ADDRESS & 351 (-64\%) & 332 (-66\%) & 304.0 & 381.0 & 77 & \num{5123311} & 53.26 & 0.15 \\
        ORIGIN & 331 (-66\%) & 316 (-68\%) & 302.0 & 341.0 & 39 & \num{78807} & 32.75 & 0.10 \\
        CALLER & 328 (-67\%) & 310 (-69\%) & 263.0 & 355.0 & 92 & \num{47434902} & 60.05 & 0.18 \\
        COINBASE & 326 (-67\%) & 311 (-68\%) & 300.0 & 331.0 & 31 & \num{58315} & 28.33 & 0.09 \\
        CALLDATALOAD & 237 (-76\%) & 214 (-78\%) & 172.0 & 269.0 & 97 & \num{68615078} & 60.82 & 0.26 \\
        MLOAD & 179 (-82\%) & 161 (-83\%) & 157.0 & 184.0 & 27 & \num{248684463} & 22.57 & 0.13 \\
        MULMOD & 176 (-82\%) & 9 (-99\%) & 9.0 & 55.5 & 46 & 3,279 & 167.45 & 0.95 \\
        SSTORE & 144 (-85\%) & 0 (-100\%) & 0.0 & 0.0 & 0 & \num{42213319} & 144.87 & 1.00 \\
        MSTORE & 143 (-85\%) & 126 (-87\%) & 113.0 & 159.0 & 46 & \num{364731080} & 38.45 & 0.27 \\
        SUICIDE & 105 (-89\%) & 0 (-100\%) & 0.0 & 0.0 & 0 & 1,123 & 105.40 & 1.00 \\
        AND & 102 (-89\%) & 26 (-97\%) & 19.0 & 40.0 & 21 & \num{285047447} & 83.26 & 0.81 \\
        CALLDATASIZE & 100 (-89\%) & 48 (-95\%) & 41.0 & 102.0 & 61 & \num{26912010} & 61.58 & 0.61 \\
        \bottomrule
        \multicolumn{7}{l}{
          \textdagger\ represents opcodes that belong in the
          \texttt{external} category which require read/write access to the state
          trie.
        }
    \end{tabular}
    \label{tab:loula-table}
\end{table*}

\begin{figure}[htbp]
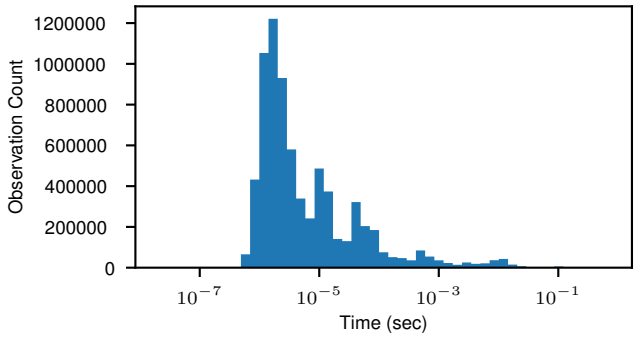

  \InputIfFileExists{plots/LTTNG-LOOKUPTIMES.pgf}{}{\textbf{!! Missing Graphics !!}}
  \caption{Distribution of Trie DB lookup times when \texttt{EXTCODESIZE} or
  \texttt{BALANCE} was executed in the EVM. The sampled time measurements
  include both cache hit and miss lookups. The distribution of time measurements
  show variability in database lookup times with a median of
  2.8us\label{fig:db-lookup-times}.}
\end{figure}

\Cref{fig:db-lookup-times} is a diagram illustrating the measured execution
time of the Trie DB lookup function used by state accessing EVM opcodes like
\texttt{EXTCODESIZE} and \texttt{BALANCE}.  The sampled interval is between
\num{3917699} and \num{4156785} on Machine B (total \num{13745782} transactions). These time
measurements were derived from traces were collected using code instrumented
with the \texttt{lttng} tracing framework.  Due to limited privileges on
Machine A and disk capacity limitations on Machine B, we could record these
dtailed traces only on Machine B and only for a limited workload.
We discovered that the
majority of time spent is focused on database operations associated with LevelDB
instead of transaction execution.  Duplication of EVM tracing show that the
\texttt{lttng} traces include approximately 100ns worth of overhead time.

%% file: plots/LTTNG-LOOKUPTIMES.pgf
\begingroup%
\makeatletter%
\begin{pgfpicture}%
\pgfpathrectangle{\pgfpointorigin}{\pgfqpoint{3.487079in}{1.941244in}}%
\pgfusepath{use as bounding box, clip}%
\begin{pgfscope}%
\pgfsetbuttcap%
\pgfsetmiterjoin%
\definecolor{currentfill}{rgb}{1.000000,1.000000,1.000000}%
\pgfsetfillcolor{currentfill}%
\pgfsetlinewidth{0.000000pt}%
\definecolor{currentstroke}{rgb}{1.000000,1.000000,1.000000}%
\pgfsetstrokecolor{currentstroke}%
\pgfsetdash{}{0pt}%
\pgfpathmoveto{\pgfqpoint{0.000000in}{0.000000in}}%
\pgfpathlineto{\pgfqpoint{3.487079in}{0.000000in}}%
\pgfpathlineto{\pgfqpoint{3.487079in}{1.941244in}}%
\pgfpathlineto{\pgfqpoint{0.000000in}{1.941244in}}%
\pgfpathclose%
\pgfusepath{fill}%
\end{pgfscope}%
\begin{pgfscope}%
\pgfsetbuttcap%
\pgfsetmiterjoin%
\definecolor{currentfill}{rgb}{1.000000,1.000000,1.000000}%
\pgfsetfillcolor{currentfill}%
\pgfsetlinewidth{0.000000pt}%
\definecolor{currentstroke}{rgb}{0.000000,0.000000,0.000000}%
\pgfsetstrokecolor{currentstroke}%
\pgfsetstrokeopacity{0.000000}%
\pgfsetdash{}{0pt}%
\pgfpathmoveto{\pgfqpoint{0.779857in}{0.442022in}}%
\pgfpathlineto{\pgfqpoint{3.352079in}{0.442022in}}%
\pgfpathlineto{\pgfqpoint{3.352079in}{1.806244in}}%
\pgfpathlineto{\pgfqpoint{0.779857in}{1.806244in}}%
\pgfpathclose%
\pgfusepath{fill}%
\end{pgfscope}%
\begin{pgfscope}%
\pgfpathrectangle{\pgfqpoint{0.779857in}{0.442022in}}{\pgfqpoint{2.572222in}{1.364222in}}%
\pgfusepath{clip}%
\pgfsetbuttcap%
\pgfsetmiterjoin%
\definecolor{currentfill}{rgb}{0.121569,0.466667,0.705882}%
\pgfsetfillcolor{currentfill}%
\pgfsetlinewidth{0.000000pt}%
\definecolor{currentstroke}{rgb}{0.000000,0.000000,0.000000}%
\pgfsetstrokecolor{currentstroke}%
\pgfsetstrokeopacity{0.000000}%
\pgfsetdash{}{0pt}%
\pgfpathmoveto{\pgfqpoint{0.896776in}{0.442022in}}%
\pgfpathlineto{\pgfqpoint{0.896776in}{0.442022in}}%
\pgfpathlineto{\pgfqpoint{0.944498in}{0.442022in}}%
\pgfpathlineto{\pgfqpoint{0.944498in}{0.442023in}}%
\pgfpathlineto{\pgfqpoint{0.992220in}{0.442023in}}%
\pgfpathlineto{\pgfqpoint{0.992220in}{0.442023in}}%
\pgfpathlineto{\pgfqpoint{1.039942in}{0.442023in}}%
\pgfpathlineto{\pgfqpoint{1.039942in}{0.442023in}}%
\pgfpathlineto{\pgfqpoint{1.087665in}{0.442023in}}%
\pgfpathlineto{\pgfqpoint{1.087665in}{0.442024in}}%
\pgfpathlineto{\pgfqpoint{1.135387in}{0.442024in}}%
\pgfpathlineto{\pgfqpoint{1.135387in}{0.442025in}}%
\pgfpathlineto{\pgfqpoint{1.183109in}{0.442025in}}%
\pgfpathlineto{\pgfqpoint{1.183109in}{0.442025in}}%
\pgfpathlineto{\pgfqpoint{1.230831in}{0.442025in}}%
\pgfpathlineto{\pgfqpoint{1.230831in}{0.442029in}}%
\pgfpathlineto{\pgfqpoint{1.278553in}{0.442029in}}%
\pgfpathlineto{\pgfqpoint{1.278553in}{0.443577in}}%
\pgfpathlineto{\pgfqpoint{1.326275in}{0.443577in}}%
\pgfpathlineto{\pgfqpoint{1.326275in}{0.511737in}}%
\pgfpathlineto{\pgfqpoint{1.373997in}{0.511737in}}%
\pgfpathlineto{\pgfqpoint{1.373997in}{0.901877in}}%
\pgfpathlineto{\pgfqpoint{1.421719in}{0.901877in}}%
\pgfpathlineto{\pgfqpoint{1.421719in}{1.562806in}}%
\pgfpathlineto{\pgfqpoint{1.469442in}{1.562806in}}%
\pgfpathlineto{\pgfqpoint{1.469442in}{1.741281in}}%
\pgfpathlineto{\pgfqpoint{1.517164in}{1.741281in}}%
\pgfpathlineto{\pgfqpoint{1.517164in}{1.431940in}}%
\pgfpathlineto{\pgfqpoint{1.564886in}{1.431940in}}%
\pgfpathlineto{\pgfqpoint{1.564886in}{1.059376in}}%
\pgfpathlineto{\pgfqpoint{1.612608in}{1.059376in}}%
\pgfpathlineto{\pgfqpoint{1.612608in}{0.802891in}}%
\pgfpathlineto{\pgfqpoint{1.660330in}{0.802891in}}%
\pgfpathlineto{\pgfqpoint{1.660330in}{0.699194in}}%
\pgfpathlineto{\pgfqpoint{1.708052in}{0.699194in}}%
\pgfpathlineto{\pgfqpoint{1.708052in}{0.959593in}}%
\pgfpathlineto{\pgfqpoint{1.755774in}{0.959593in}}%
\pgfpathlineto{\pgfqpoint{1.755774in}{0.839684in}}%
\pgfpathlineto{\pgfqpoint{1.803496in}{0.839684in}}%
\pgfpathlineto{\pgfqpoint{1.803496in}{0.591967in}}%
\pgfpathlineto{\pgfqpoint{1.851219in}{0.591967in}}%
\pgfpathlineto{\pgfqpoint{1.851219in}{0.580520in}}%
\pgfpathlineto{\pgfqpoint{1.898941in}{0.580520in}}%
\pgfpathlineto{\pgfqpoint{1.898941in}{0.784468in}}%
\pgfpathlineto{\pgfqpoint{1.946663in}{0.784468in}}%
\pgfpathlineto{\pgfqpoint{1.946663in}{0.659072in}}%
\pgfpathlineto{\pgfqpoint{1.994385in}{0.659072in}}%
\pgfpathlineto{\pgfqpoint{1.994385in}{0.638431in}}%
\pgfpathlineto{\pgfqpoint{2.042107in}{0.638431in}}%
\pgfpathlineto{\pgfqpoint{2.042107in}{0.522382in}}%
\pgfpathlineto{\pgfqpoint{2.089829in}{0.522382in}}%
\pgfpathlineto{\pgfqpoint{2.089829in}{0.496954in}}%
\pgfpathlineto{\pgfqpoint{2.137551in}{0.496954in}}%
\pgfpathlineto{\pgfqpoint{2.137551in}{0.491946in}}%
\pgfpathlineto{\pgfqpoint{2.185273in}{0.491946in}}%
\pgfpathlineto{\pgfqpoint{2.185273in}{0.480435in}}%
\pgfpathlineto{\pgfqpoint{2.232995in}{0.480435in}}%
\pgfpathlineto{\pgfqpoint{2.232995in}{0.531946in}}%
\pgfpathlineto{\pgfqpoint{2.280718in}{0.531946in}}%
\pgfpathlineto{\pgfqpoint{2.280718in}{0.500027in}}%
\pgfpathlineto{\pgfqpoint{2.328440in}{0.500027in}}%
\pgfpathlineto{\pgfqpoint{2.328440in}{0.480382in}}%
\pgfpathlineto{\pgfqpoint{2.376162in}{0.480382in}}%
\pgfpathlineto{\pgfqpoint{2.376162in}{0.466434in}}%
\pgfpathlineto{\pgfqpoint{2.423884in}{0.466434in}}%
\pgfpathlineto{\pgfqpoint{2.423884in}{0.457170in}}%
\pgfpathlineto{\pgfqpoint{2.471606in}{0.457170in}}%
\pgfpathlineto{\pgfqpoint{2.471606in}{0.469181in}}%
\pgfpathlineto{\pgfqpoint{2.519328in}{0.469181in}}%
\pgfpathlineto{\pgfqpoint{2.519328in}{0.462152in}}%
\pgfpathlineto{\pgfqpoint{2.567050in}{0.462152in}}%
\pgfpathlineto{\pgfqpoint{2.567050in}{0.464748in}}%
\pgfpathlineto{\pgfqpoint{2.614772in}{0.464748in}}%
\pgfpathlineto{\pgfqpoint{2.614772in}{0.480979in}}%
\pgfpathlineto{\pgfqpoint{2.662495in}{0.480979in}}%
\pgfpathlineto{\pgfqpoint{2.662495in}{0.487600in}}%
\pgfpathlineto{\pgfqpoint{2.710217in}{0.487600in}}%
\pgfpathlineto{\pgfqpoint{2.710217in}{0.458134in}}%
\pgfpathlineto{\pgfqpoint{2.757939in}{0.458134in}}%
\pgfpathlineto{\pgfqpoint{2.757939in}{0.449636in}}%
\pgfpathlineto{\pgfqpoint{2.805661in}{0.449636in}}%
\pgfpathlineto{\pgfqpoint{2.805661in}{0.443211in}}%
\pgfpathlineto{\pgfqpoint{2.853383in}{0.443211in}}%
\pgfpathlineto{\pgfqpoint{2.853383in}{0.442461in}}%
\pgfpathlineto{\pgfqpoint{2.901105in}{0.442461in}}%
\pgfpathlineto{\pgfqpoint{2.901105in}{0.444103in}}%
\pgfpathlineto{\pgfqpoint{2.948827in}{0.444103in}}%
\pgfpathlineto{\pgfqpoint{2.948827in}{0.448380in}}%
\pgfpathlineto{\pgfqpoint{2.996549in}{0.448380in}}%
\pgfpathlineto{\pgfqpoint{2.996549in}{0.442524in}}%
\pgfpathlineto{\pgfqpoint{3.044271in}{0.442524in}}%
\pgfpathlineto{\pgfqpoint{3.044271in}{0.442281in}}%
\pgfpathlineto{\pgfqpoint{3.091994in}{0.442281in}}%
\pgfpathlineto{\pgfqpoint{3.091994in}{0.442195in}}%
\pgfpathlineto{\pgfqpoint{3.139716in}{0.442195in}}%
\pgfpathlineto{\pgfqpoint{3.139716in}{0.442176in}}%
\pgfpathlineto{\pgfqpoint{3.187438in}{0.442176in}}%
\pgfpathlineto{\pgfqpoint{3.187438in}{0.442068in}}%
\pgfpathlineto{\pgfqpoint{3.235160in}{0.442068in}}%
\pgfpathlineto{\pgfqpoint{3.235160in}{0.442022in}}%
\pgfpathlineto{\pgfqpoint{3.187438in}{0.442022in}}%
\pgfpathlineto{\pgfqpoint{3.187438in}{0.442022in}}%
\pgfpathlineto{\pgfqpoint{3.139716in}{0.442022in}}%
\pgfpathlineto{\pgfqpoint{3.139716in}{0.442022in}}%
\pgfpathlineto{\pgfqpoint{3.091994in}{0.442022in}}%
\pgfpathlineto{\pgfqpoint{3.091994in}{0.442022in}}%
\pgfpathlineto{\pgfqpoint{3.044271in}{0.442022in}}%
\pgfpathlineto{\pgfqpoint{3.044271in}{0.442022in}}%
\pgfpathlineto{\pgfqpoint{2.996549in}{0.442022in}}%
\pgfpathlineto{\pgfqpoint{2.996549in}{0.442022in}}%
\pgfpathlineto{\pgfqpoint{2.948827in}{0.442022in}}%
\pgfpathlineto{\pgfqpoint{2.948827in}{0.442022in}}%
\pgfpathlineto{\pgfqpoint{2.901105in}{0.442022in}}%
\pgfpathlineto{\pgfqpoint{2.901105in}{0.442022in}}%
\pgfpathlineto{\pgfqpoint{2.853383in}{0.442022in}}%
\pgfpathlineto{\pgfqpoint{2.853383in}{0.442022in}}%
\pgfpathlineto{\pgfqpoint{2.805661in}{0.442022in}}%
\pgfpathlineto{\pgfqpoint{2.805661in}{0.442022in}}%
\pgfpathlineto{\pgfqpoint{2.757939in}{0.442022in}}%
\pgfpathlineto{\pgfqpoint{2.757939in}{0.442022in}}%
\pgfpathlineto{\pgfqpoint{2.710217in}{0.442022in}}%
\pgfpathlineto{\pgfqpoint{2.710217in}{0.442022in}}%
\pgfpathlineto{\pgfqpoint{2.662495in}{0.442022in}}%
\pgfpathlineto{\pgfqpoint{2.662495in}{0.442022in}}%
\pgfpathlineto{\pgfqpoint{2.614772in}{0.442022in}}%
\pgfpathlineto{\pgfqpoint{2.614772in}{0.442022in}}%
\pgfpathlineto{\pgfqpoint{2.567050in}{0.442022in}}%
\pgfpathlineto{\pgfqpoint{2.567050in}{0.442022in}}%
\pgfpathlineto{\pgfqpoint{2.519328in}{0.442022in}}%
\pgfpathlineto{\pgfqpoint{2.519328in}{0.442022in}}%
\pgfpathlineto{\pgfqpoint{2.471606in}{0.442022in}}%
\pgfpathlineto{\pgfqpoint{2.471606in}{0.442022in}}%
\pgfpathlineto{\pgfqpoint{2.423884in}{0.442022in}}%
\pgfpathlineto{\pgfqpoint{2.423884in}{0.442022in}}%
\pgfpathlineto{\pgfqpoint{2.376162in}{0.442022in}}%
\pgfpathlineto{\pgfqpoint{2.376162in}{0.442022in}}%
\pgfpathlineto{\pgfqpoint{2.328440in}{0.442022in}}%
\pgfpathlineto{\pgfqpoint{2.328440in}{0.442022in}}%
\pgfpathlineto{\pgfqpoint{2.280718in}{0.442022in}}%
\pgfpathlineto{\pgfqpoint{2.280718in}{0.442022in}}%
\pgfpathlineto{\pgfqpoint{2.232995in}{0.442022in}}%
\pgfpathlineto{\pgfqpoint{2.232995in}{0.442022in}}%
\pgfpathlineto{\pgfqpoint{2.185273in}{0.442022in}}%
\pgfpathlineto{\pgfqpoint{2.185273in}{0.442022in}}%
\pgfpathlineto{\pgfqpoint{2.137551in}{0.442022in}}%
\pgfpathlineto{\pgfqpoint{2.137551in}{0.442022in}}%
\pgfpathlineto{\pgfqpoint{2.089829in}{0.442022in}}%
\pgfpathlineto{\pgfqpoint{2.089829in}{0.442022in}}%
\pgfpathlineto{\pgfqpoint{2.042107in}{0.442022in}}%
\pgfpathlineto{\pgfqpoint{2.042107in}{0.442022in}}%
\pgfpathlineto{\pgfqpoint{1.994385in}{0.442022in}}%
\pgfpathlineto{\pgfqpoint{1.994385in}{0.442022in}}%
\pgfpathlineto{\pgfqpoint{1.946663in}{0.442022in}}%
\pgfpathlineto{\pgfqpoint{1.946663in}{0.442022in}}%
\pgfpathlineto{\pgfqpoint{1.898941in}{0.442022in}}%
\pgfpathlineto{\pgfqpoint{1.898941in}{0.442022in}}%
\pgfpathlineto{\pgfqpoint{1.851219in}{0.442022in}}%
\pgfpathlineto{\pgfqpoint{1.851219in}{0.442022in}}%
\pgfpathlineto{\pgfqpoint{1.803496in}{0.442022in}}%
\pgfpathlineto{\pgfqpoint{1.803496in}{0.442022in}}%
\pgfpathlineto{\pgfqpoint{1.755774in}{0.442022in}}%
\pgfpathlineto{\pgfqpoint{1.755774in}{0.442022in}}%
\pgfpathlineto{\pgfqpoint{1.708052in}{0.442022in}}%
\pgfpathlineto{\pgfqpoint{1.708052in}{0.442022in}}%
\pgfpathlineto{\pgfqpoint{1.660330in}{0.442022in}}%
\pgfpathlineto{\pgfqpoint{1.660330in}{0.442022in}}%
\pgfpathlineto{\pgfqpoint{1.612608in}{0.442022in}}%
\pgfpathlineto{\pgfqpoint{1.612608in}{0.442022in}}%
\pgfpathlineto{\pgfqpoint{1.564886in}{0.442022in}}%
\pgfpathlineto{\pgfqpoint{1.564886in}{0.442022in}}%
\pgfpathlineto{\pgfqpoint{1.517164in}{0.442022in}}%
\pgfpathlineto{\pgfqpoint{1.517164in}{0.442022in}}%
\pgfpathlineto{\pgfqpoint{1.469442in}{0.442022in}}%
\pgfpathlineto{\pgfqpoint{1.469442in}{0.442022in}}%
\pgfpathlineto{\pgfqpoint{1.421719in}{0.442022in}}%
\pgfpathlineto{\pgfqpoint{1.421719in}{0.442022in}}%
\pgfpathlineto{\pgfqpoint{1.373997in}{0.442022in}}%
\pgfpathlineto{\pgfqpoint{1.373997in}{0.442022in}}%
\pgfpathlineto{\pgfqpoint{1.326275in}{0.442022in}}%
\pgfpathlineto{\pgfqpoint{1.326275in}{0.442022in}}%
\pgfpathlineto{\pgfqpoint{1.278553in}{0.442022in}}%
\pgfpathlineto{\pgfqpoint{1.278553in}{0.442022in}}%
\pgfpathlineto{\pgfqpoint{1.230831in}{0.442022in}}%
\pgfpathlineto{\pgfqpoint{1.230831in}{0.442022in}}%
\pgfpathlineto{\pgfqpoint{1.183109in}{0.442022in}}%
\pgfpathlineto{\pgfqpoint{1.183109in}{0.442022in}}%
\pgfpathlineto{\pgfqpoint{1.135387in}{0.442022in}}%
\pgfpathlineto{\pgfqpoint{1.135387in}{0.442022in}}%
\pgfpathlineto{\pgfqpoint{1.087665in}{0.442022in}}%
\pgfpathlineto{\pgfqpoint{1.087665in}{0.442022in}}%
\pgfpathlineto{\pgfqpoint{1.039942in}{0.442022in}}%
\pgfpathlineto{\pgfqpoint{1.039942in}{0.442022in}}%
\pgfpathlineto{\pgfqpoint{0.992220in}{0.442022in}}%
\pgfpathlineto{\pgfqpoint{0.992220in}{0.442022in}}%
\pgfpathlineto{\pgfqpoint{0.944498in}{0.442022in}}%
\pgfpathlineto{\pgfqpoint{0.944498in}{0.442022in}}%
\pgfpathclose%
\pgfusepath{fill}%
\end{pgfscope}%
\begin{pgfscope}%
\pgfsetbuttcap%
\pgfsetroundjoin%
\definecolor{currentfill}{rgb}{0.000000,0.000000,0.000000}%
\pgfsetfillcolor{currentfill}%
\pgfsetlinewidth{0.803000pt}%
\definecolor{currentstroke}{rgb}{0.000000,0.000000,0.000000}%
\pgfsetstrokecolor{currentstroke}%
\pgfsetdash{}{0pt}%
\pgfsys@defobject{currentmarker}{\pgfqpoint{0.000000in}{-0.048611in}}{\pgfqpoint{0.000000in}{0.000000in}}{%
\pgfpathmoveto{\pgfqpoint{0.000000in}{0.000000in}}%
\pgfpathlineto{\pgfqpoint{0.000000in}{-0.048611in}}%
\pgfusepath{stroke,fill}%
}%
\begin{pgfscope}%
\pgfsys@transformshift{1.113151in}{0.442022in}%
\pgfsys@useobject{currentmarker}{}%
\end{pgfscope}%
\end{pgfscope}%
\begin{pgfscope}%
\definecolor{textcolor}{rgb}{0.000000,0.000000,0.000000}%
\pgfsetstrokecolor{textcolor}%
\pgfsetfillcolor{textcolor}%
\pgftext[x=1.113151in,y=0.344799in,,top]{\color{textcolor}\sffamily\fontsize{7.000000}{8.400000}\selectfont \(\displaystyle {10^{-7}}\)}%
\end{pgfscope}%
\begin{pgfscope}%
\pgfsetbuttcap%
\pgfsetroundjoin%
\definecolor{currentfill}{rgb}{0.000000,0.000000,0.000000}%
\pgfsetfillcolor{currentfill}%
\pgfsetlinewidth{0.803000pt}%
\definecolor{currentstroke}{rgb}{0.000000,0.000000,0.000000}%
\pgfsetstrokecolor{currentstroke}%
\pgfsetdash{}{0pt}%
\pgfsys@defobject{currentmarker}{\pgfqpoint{0.000000in}{-0.048611in}}{\pgfqpoint{0.000000in}{0.000000in}}{%
\pgfpathmoveto{\pgfqpoint{0.000000in}{0.000000in}}%
\pgfpathlineto{\pgfqpoint{0.000000in}{-0.048611in}}%
\pgfusepath{stroke,fill}%
}%
\begin{pgfscope}%
\pgfsys@transformshift{1.732278in}{0.442022in}%
\pgfsys@useobject{currentmarker}{}%
\end{pgfscope}%
\end{pgfscope}%
\begin{pgfscope}%
\definecolor{textcolor}{rgb}{0.000000,0.000000,0.000000}%
\pgfsetstrokecolor{textcolor}%
\pgfsetfillcolor{textcolor}%
\pgftext[x=1.732278in,y=0.344799in,,top]{\color{textcolor}\sffamily\fontsize{7.000000}{8.400000}\selectfont \(\displaystyle {10^{-5}}\)}%
\end{pgfscope}%
\begin{pgfscope}%
\pgfsetbuttcap%
\pgfsetroundjoin%
\definecolor{currentfill}{rgb}{0.000000,0.000000,0.000000}%
\pgfsetfillcolor{currentfill}%
\pgfsetlinewidth{0.803000pt}%
\definecolor{currentstroke}{rgb}{0.000000,0.000000,0.000000}%
\pgfsetstrokecolor{currentstroke}%
\pgfsetdash{}{0pt}%
\pgfsys@defobject{currentmarker}{\pgfqpoint{0.000000in}{-0.048611in}}{\pgfqpoint{0.000000in}{0.000000in}}{%
\pgfpathmoveto{\pgfqpoint{0.000000in}{0.000000in}}%
\pgfpathlineto{\pgfqpoint{0.000000in}{-0.048611in}}%
\pgfusepath{stroke,fill}%
}%
\begin{pgfscope}%
\pgfsys@transformshift{2.351404in}{0.442022in}%
\pgfsys@useobject{currentmarker}{}%
\end{pgfscope}%
\end{pgfscope}%
\begin{pgfscope}%
\definecolor{textcolor}{rgb}{0.000000,0.000000,0.000000}%
\pgfsetstrokecolor{textcolor}%
\pgfsetfillcolor{textcolor}%
\pgftext[x=2.351404in,y=0.344799in,,top]{\color{textcolor}\sffamily\fontsize{7.000000}{8.400000}\selectfont \(\displaystyle {10^{-3}}\)}%
\end{pgfscope}%
\begin{pgfscope}%
\pgfsetbuttcap%
\pgfsetroundjoin%
\definecolor{currentfill}{rgb}{0.000000,0.000000,0.000000}%
\pgfsetfillcolor{currentfill}%
\pgfsetlinewidth{0.803000pt}%
\definecolor{currentstroke}{rgb}{0.000000,0.000000,0.000000}%
\pgfsetstrokecolor{currentstroke}%
\pgfsetdash{}{0pt}%
\pgfsys@defobject{currentmarker}{\pgfqpoint{0.000000in}{-0.048611in}}{\pgfqpoint{0.000000in}{0.000000in}}{%
\pgfpathmoveto{\pgfqpoint{0.000000in}{0.000000in}}%
\pgfpathlineto{\pgfqpoint{0.000000in}{-0.048611in}}%
\pgfusepath{stroke,fill}%
}%
\begin{pgfscope}%
\pgfsys@transformshift{2.970530in}{0.442022in}%
\pgfsys@useobject{currentmarker}{}%
\end{pgfscope}%
\end{pgfscope}%
\begin{pgfscope}%
\definecolor{textcolor}{rgb}{0.000000,0.000000,0.000000}%
\pgfsetstrokecolor{textcolor}%
\pgfsetfillcolor{textcolor}%
\pgftext[x=2.970530in,y=0.344799in,,top]{\color{textcolor}\sffamily\fontsize{7.000000}{8.400000}\selectfont \(\displaystyle {10^{-1}}\)}%
\end{pgfscope}%
\begin{pgfscope}%
\definecolor{textcolor}{rgb}{0.000000,0.000000,0.000000}%
\pgfsetstrokecolor{textcolor}%
\pgfsetfillcolor{textcolor}%
\pgftext[x=2.065968in,y=0.194089in,,top]{\color{textcolor}\sffamily\fontsize{7.000000}{8.400000}\selectfont Time (sec)}%
\end{pgfscope}%
\begin{pgfscope}%
\pgfsetbuttcap%
\pgfsetroundjoin%
\definecolor{currentfill}{rgb}{0.000000,0.000000,0.000000}%
\pgfsetfillcolor{currentfill}%
\pgfsetlinewidth{0.803000pt}%
\definecolor{currentstroke}{rgb}{0.000000,0.000000,0.000000}%
\pgfsetstrokecolor{currentstroke}%
\pgfsetdash{}{0pt}%
\pgfsys@defobject{currentmarker}{\pgfqpoint{-0.048611in}{0.000000in}}{\pgfqpoint{0.000000in}{0.000000in}}{%
\pgfpathmoveto{\pgfqpoint{0.000000in}{0.000000in}}%
\pgfpathlineto{\pgfqpoint{-0.048611in}{0.000000in}}%
\pgfusepath{stroke,fill}%
}%
\begin{pgfscope}%
\pgfsys@transformshift{0.779857in}{0.442022in}%
\pgfsys@useobject{currentmarker}{}%
\end{pgfscope}%
\end{pgfscope}%
\begin{pgfscope}%
\definecolor{textcolor}{rgb}{0.000000,0.000000,0.000000}%
\pgfsetstrokecolor{textcolor}%
\pgfsetfillcolor{textcolor}%
\pgftext[x=0.620779in,y=0.405088in,left,base]{\color{textcolor}\sffamily\fontsize{7.000000}{8.400000}\selectfont 0}%
\end{pgfscope}%
\begin{pgfscope}%
\pgfsetbuttcap%
\pgfsetroundjoin%
\definecolor{currentfill}{rgb}{0.000000,0.000000,0.000000}%
\pgfsetfillcolor{currentfill}%
\pgfsetlinewidth{0.803000pt}%
\definecolor{currentstroke}{rgb}{0.000000,0.000000,0.000000}%
\pgfsetstrokecolor{currentstroke}%
\pgfsetdash{}{0pt}%
\pgfsys@defobject{currentmarker}{\pgfqpoint{-0.048611in}{0.000000in}}{\pgfqpoint{0.000000in}{0.000000in}}{%
\pgfpathmoveto{\pgfqpoint{0.000000in}{0.000000in}}%
\pgfpathlineto{\pgfqpoint{-0.048611in}{0.000000in}}%
\pgfusepath{stroke,fill}%
}%
\begin{pgfscope}%
\pgfsys@transformshift{0.779857in}{0.654810in}%
\pgfsys@useobject{currentmarker}{}%
\end{pgfscope}%
\end{pgfscope}%
\begin{pgfscope}%
\definecolor{textcolor}{rgb}{0.000000,0.000000,0.000000}%
\pgfsetstrokecolor{textcolor}%
\pgfsetfillcolor{textcolor}%
\pgftext[x=0.311500in,y=0.617877in,left,base]{\color{textcolor}\sffamily\fontsize{7.000000}{8.400000}\selectfont 200000}%
\end{pgfscope}%
\begin{pgfscope}%
\pgfsetbuttcap%
\pgfsetroundjoin%
\definecolor{currentfill}{rgb}{0.000000,0.000000,0.000000}%
\pgfsetfillcolor{currentfill}%
\pgfsetlinewidth{0.803000pt}%
\definecolor{currentstroke}{rgb}{0.000000,0.000000,0.000000}%
\pgfsetstrokecolor{currentstroke}%
\pgfsetdash{}{0pt}%
\pgfsys@defobject{currentmarker}{\pgfqpoint{-0.048611in}{0.000000in}}{\pgfqpoint{0.000000in}{0.000000in}}{%
\pgfpathmoveto{\pgfqpoint{0.000000in}{0.000000in}}%
\pgfpathlineto{\pgfqpoint{-0.048611in}{0.000000in}}%
\pgfusepath{stroke,fill}%
}%
\begin{pgfscope}%
\pgfsys@transformshift{0.779857in}{0.867598in}%
\pgfsys@useobject{currentmarker}{}%
\end{pgfscope}%
\end{pgfscope}%
\begin{pgfscope}%
\definecolor{textcolor}{rgb}{0.000000,0.000000,0.000000}%
\pgfsetstrokecolor{textcolor}%
\pgfsetfillcolor{textcolor}%
\pgftext[x=0.311500in,y=0.830665in,left,base]{\color{textcolor}\sffamily\fontsize{7.000000}{8.400000}\selectfont 400000}%
\end{pgfscope}%
\begin{pgfscope}%
\pgfsetbuttcap%
\pgfsetroundjoin%
\definecolor{currentfill}{rgb}{0.000000,0.000000,0.000000}%
\pgfsetfillcolor{currentfill}%
\pgfsetlinewidth{0.803000pt}%
\definecolor{currentstroke}{rgb}{0.000000,0.000000,0.000000}%
\pgfsetstrokecolor{currentstroke}%
\pgfsetdash{}{0pt}%
\pgfsys@defobject{currentmarker}{\pgfqpoint{-0.048611in}{0.000000in}}{\pgfqpoint{0.000000in}{0.000000in}}{%
\pgfpathmoveto{\pgfqpoint{0.000000in}{0.000000in}}%
\pgfpathlineto{\pgfqpoint{-0.048611in}{0.000000in}}%
\pgfusepath{stroke,fill}%
}%
\begin{pgfscope}%
\pgfsys@transformshift{0.779857in}{1.080387in}%
\pgfsys@useobject{currentmarker}{}%
\end{pgfscope}%
\end{pgfscope}%
\begin{pgfscope}%
\definecolor{textcolor}{rgb}{0.000000,0.000000,0.000000}%
\pgfsetstrokecolor{textcolor}%
\pgfsetfillcolor{textcolor}%
\pgftext[x=0.311500in,y=1.043454in,left,base]{\color{textcolor}\sffamily\fontsize{7.000000}{8.400000}\selectfont 600000}%
\end{pgfscope}%
\begin{pgfscope}%
\pgfsetbuttcap%
\pgfsetroundjoin%
\definecolor{currentfill}{rgb}{0.000000,0.000000,0.000000}%
\pgfsetfillcolor{currentfill}%
\pgfsetlinewidth{0.803000pt}%
\definecolor{currentstroke}{rgb}{0.000000,0.000000,0.000000}%
\pgfsetstrokecolor{currentstroke}%
\pgfsetdash{}{0pt}%
\pgfsys@defobject{currentmarker}{\pgfqpoint{-0.048611in}{0.000000in}}{\pgfqpoint{0.000000in}{0.000000in}}{%
\pgfpathmoveto{\pgfqpoint{0.000000in}{0.000000in}}%
\pgfpathlineto{\pgfqpoint{-0.048611in}{0.000000in}}%
\pgfusepath{stroke,fill}%
}%
\begin{pgfscope}%
\pgfsys@transformshift{0.779857in}{1.293175in}%
\pgfsys@useobject{currentmarker}{}%
\end{pgfscope}%
\end{pgfscope}%
\begin{pgfscope}%
\definecolor{textcolor}{rgb}{0.000000,0.000000,0.000000}%
\pgfsetstrokecolor{textcolor}%
\pgfsetfillcolor{textcolor}%
\pgftext[x=0.311500in,y=1.256242in,left,base]{\color{textcolor}\sffamily\fontsize{7.000000}{8.400000}\selectfont 800000}%
\end{pgfscope}%
\begin{pgfscope}%
\pgfsetbuttcap%
\pgfsetroundjoin%
\definecolor{currentfill}{rgb}{0.000000,0.000000,0.000000}%
\pgfsetfillcolor{currentfill}%
\pgfsetlinewidth{0.803000pt}%
\definecolor{currentstroke}{rgb}{0.000000,0.000000,0.000000}%
\pgfsetstrokecolor{currentstroke}%
\pgfsetdash{}{0pt}%
\pgfsys@defobject{currentmarker}{\pgfqpoint{-0.048611in}{0.000000in}}{\pgfqpoint{0.000000in}{0.000000in}}{%
\pgfpathmoveto{\pgfqpoint{0.000000in}{0.000000in}}%
\pgfpathlineto{\pgfqpoint{-0.048611in}{0.000000in}}%
\pgfusepath{stroke,fill}%
}%
\begin{pgfscope}%
\pgfsys@transformshift{0.779857in}{1.505963in}%
\pgfsys@useobject{currentmarker}{}%
\end{pgfscope}%
\end{pgfscope}%
\begin{pgfscope}%
\definecolor{textcolor}{rgb}{0.000000,0.000000,0.000000}%
\pgfsetstrokecolor{textcolor}%
\pgfsetfillcolor{textcolor}%
\pgftext[x=0.249644in,y=1.469030in,left,base]{\color{textcolor}\sffamily\fontsize{7.000000}{8.400000}\selectfont 1000000}%
\end{pgfscope}%
\begin{pgfscope}%
\pgfsetbuttcap%
\pgfsetroundjoin%
\definecolor{currentfill}{rgb}{0.000000,0.000000,0.000000}%
\pgfsetfillcolor{currentfill}%
\pgfsetlinewidth{0.803000pt}%
\definecolor{currentstroke}{rgb}{0.000000,0.000000,0.000000}%
\pgfsetstrokecolor{currentstroke}%
\pgfsetdash{}{0pt}%
\pgfsys@defobject{currentmarker}{\pgfqpoint{-0.048611in}{0.000000in}}{\pgfqpoint{0.000000in}{0.000000in}}{%
\pgfpathmoveto{\pgfqpoint{0.000000in}{0.000000in}}%
\pgfpathlineto{\pgfqpoint{-0.048611in}{0.000000in}}%
\pgfusepath{stroke,fill}%
}%
\begin{pgfscope}%
\pgfsys@transformshift{0.779857in}{1.718752in}%
\pgfsys@useobject{currentmarker}{}%
\end{pgfscope}%
\end{pgfscope}%
\begin{pgfscope}%
\definecolor{textcolor}{rgb}{0.000000,0.000000,0.000000}%
\pgfsetstrokecolor{textcolor}%
\pgfsetfillcolor{textcolor}%
\pgftext[x=0.249644in,y=1.681819in,left,base]{\color{textcolor}\sffamily\fontsize{7.000000}{8.400000}\selectfont 1200000}%
\end{pgfscope}%
\begin{pgfscope}%
\definecolor{textcolor}{rgb}{0.000000,0.000000,0.000000}%
\pgfsetstrokecolor{textcolor}%
\pgfsetfillcolor{textcolor}%
\pgftext[x=0.194089in,y=1.124133in,,bottom,rotate=90.000000]{\color{textcolor}\sffamily\fontsize{7.000000}{8.400000}\selectfont Observation Count}%
\end{pgfscope}%
\begin{pgfscope}%
\pgfsetrectcap%
\pgfsetmiterjoin%
\pgfsetlinewidth{0.803000pt}%
\definecolor{currentstroke}{rgb}{0.000000,0.000000,0.000000}%
\pgfsetstrokecolor{currentstroke}%
\pgfsetdash{}{0pt}%
\pgfpathmoveto{\pgfqpoint{0.779857in}{0.442022in}}%
\pgfpathlineto{\pgfqpoint{0.779857in}{1.806244in}}%
\pgfusepath{stroke}%
\end{pgfscope}%
\begin{pgfscope}%
\pgfsetrectcap%
\pgfsetmiterjoin%
\pgfsetlinewidth{0.803000pt}%
\definecolor{currentstroke}{rgb}{0.000000,0.000000,0.000000}%
\pgfsetstrokecolor{currentstroke}%
\pgfsetdash{}{0pt}%
\pgfpathmoveto{\pgfqpoint{3.352079in}{0.442022in}}%
\pgfpathlineto{\pgfqpoint{3.352079in}{1.806244in}}%
\pgfusepath{stroke}%
\end{pgfscope}%
\begin{pgfscope}%
\pgfsetrectcap%
\pgfsetmiterjoin%
\pgfsetlinewidth{0.803000pt}%
\definecolor{currentstroke}{rgb}{0.000000,0.000000,0.000000}%
\pgfsetstrokecolor{currentstroke}%
\pgfsetdash{}{0pt}%
\pgfpathmoveto{\pgfqpoint{0.779857in}{0.442022in}}%
\pgfpathlineto{\pgfqpoint{3.352079in}{0.442022in}}%
\pgfusepath{stroke}%
\end{pgfscope}%
\begin{pgfscope}%
\pgfsetrectcap%
\pgfsetmiterjoin%
\pgfsetlinewidth{0.803000pt}%
\definecolor{currentstroke}{rgb}{0.000000,0.000000,0.000000}%
\pgfsetstrokecolor{currentstroke}%
\pgfsetdash{}{0pt}%
\pgfpathmoveto{\pgfqpoint{0.779857in}{1.806244in}}%
\pgfpathlineto{\pgfqpoint{3.352079in}{1.806244in}}%
\pgfusepath{stroke}%
\end{pgfscope}%
\end{pgfpicture}%
\makeatother%
\endgroup%

%% file: content/discussion.tex
\section{Discussion}
From the results we infer three things.  Firstly, certain opcode execution
times have exceeded well beyond the baseline, which poses an attack vector for
denial of service.  Secondly, some opcodes have high variance in their ratio
distributions further weakening the linear correlation between execution time
and assigned gas cost.  This poses an issue as an adversary can exploit this by
authoring transactions that mimic the execution profile of high time-to-gas
ratio transactions.  Finally, our results also show that machines with different
hardware capability yield different EVM execution profiles.  We will
attempt to explain the cause of these observations in this section.

\paragraph{Denial-of-Service}
\begin{figure}[htbp]
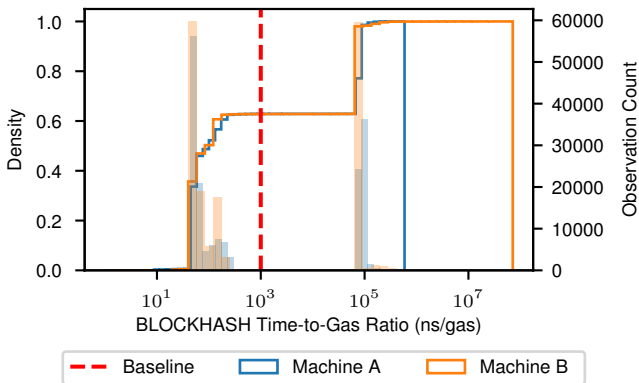

  \InputIfFileExists{plots/BLOCKHASH.pgf}{}{\textbf{!! Missing Graphics !!}}
  \caption{Histogram plot (represented by shaded region) with CDF overlay
    (represented by solid lines) showing approximately 50\% of
    \texttt{BLOCKHASH} execution instances performing worse than the baseline
    (denoted in dotted line along the vertical axis) posing a denial-of-service
  vulnerability on both machines. Overlapping regions indicate that the ratio
distribution is identical across both machines.}
\end{figure}
Underpriced gas cost for EVM opcodes pose a denial-of-service attack vector
within the Ethereum ecosystem. While under the workload, both machines
consistently shown that the \texttt{BLOCKHASH} opcode has the highest mean
time-to-gas ratio well beyond the baseline which indicate that the opcode is
severely underpriced and exploits can cause severe disruptions to all users.

The measured IQR further affirm that the opcode has been assigned with an
incorrect gas cost as the ratio distribution is extremely sparse.

This opcode is one of the three opcode that falls under the \texttt{external}
category as defined in the Ethereum yellowpaper as it requires access to the
state trie during its execution.  The main use of the opcode is to get the hash
of one of the 256 most recent complete blocks.  The main bottleneck for this
particular opcode is caused by the fact that the \texttt{BLOCKHASH} EVM
operation is implemented as a smart contract instead of a built-in routine within
the EVM. Therefore its execution requires traversing the state trie to load the
contract code into RAM for execution.  On top of that, in the reference
implementation, there is also the overhead of creating an embedded EVM instance
within the runtime of the base EVM executing the contract, incurring more
overhead.

We note that this analysis applies only to the C++ reference implementation.
Other users might, for instance, run custom, private implementations in which the
\texttt{BLOCKHASH} opcode is implemented more efficiently, thereby avoiding
this potential denial-of-service vector.

We note further that an impending change~\cite{aleth2018blockhashpr} to the gas
cost for \texttt{BLOCKHASH} is scheduled to take place in the Constantinople
hard-fork where the gas cost is raised from 20 to 800. Based on our results,
this change would put its mean ratio below the baseline (1,000), which would fall
from \num{35000} to just 875.

\paragraph{Gas Cost Misalignment}

Risk of denial of service aside, our measurements also show that on both
experimental platforms, there are other EVM \texttt{external}-class opcodes (\ie
\texttt{BALANCE}, \texttt{SLOAD} and etc) that have relatively higher
variability in their time-to-gas ratio in comparison to other standard opcodes.
While not directly a vulnerability, higher variability in ratio distribution
make performance profiles unpredictable and opens up the possibility of
transactions being executed at the expense of an unfairly high computational
cost.  This variability is generally caused by the fact that the client software
must either load state trie objects from disk or a memory-resident cache.  Due
to the varying latencies posed by memory hierarchy in computing hardware,
reading from a SSD disk can incur a 10 to 16 time penalty; on a HDD disk, this
penalty could be 1,000 times larger.  The nondeterminism of objects being resident
in memory is the cause of the variability, which was further supported by our
\texttt{lttng} traces. The traces show that accessing a cached DB lookup is on
average 100 times faster than it takes to process a DB lookup that requires disk
access.  From our measurements, we observed that on average, executing the
\texttt{BALANCE} opcode requires 7 DB lookups, of which 1 or 2
will require disk access.

Most if not all Ethereum implementations use LevelDB as backend
storage by default.  LevelDB is a minimal key-value store database that
provides an ordered mapping from string keys to string values.  The associated
overhead of looking up state during transaction execution is a consequence of
how the state trie is stored internally within LevelDB.  Recall that the state
trie is an implementation of a Merkle Patricia Trie composed of three types of
value nodes (\ie extension node, value node and index node). The traversal of
the trie for any particular given value node has a theoretical constant lookup
time which is the size of the query key.  In the case of looking up an account
balance, since Ethereum accounts are identified by keys of 20 bytes (40 hex
characters) in length, the maximum number of traversal steps required will be
10.  If there exist multiple value nodes that share
common prefixes in their keys then this number of lookup steps can be
reduced, by utilising
extension nodes (explained below).  During a lookup the state root index is first
retrieved, which resolves to an index node. The first element in the key is then
used to retrieve subsequent index nodes or extension nodes.  Extension nodes
allow fast-track traversal along the state-trie reducing the need to traverse
the full key length and can only be created when multiple node values
share common substrings in their keys.  Given the design of how state objects
are retrieved, it is unsurprising to find that
EVM opcodes that require state trie access exhibit high variability
in their Time-to-Gas
ratios.

\begin{figure}[h!]
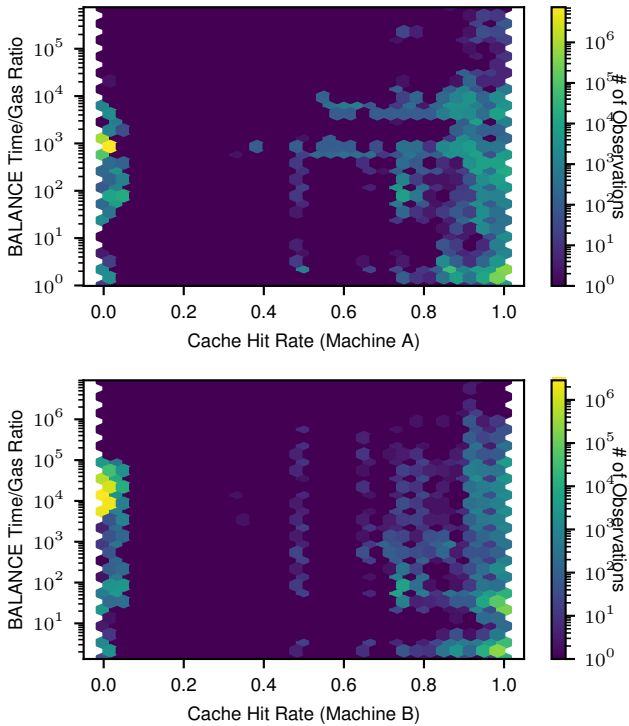

    \InputIfFileExists{plots/ACCOUNT-CACHERATE.pgf}{}{\textbf{Image File does not exist!}}
    \caption{\small{\label{fig:account-call-hitrate}
        Hexbin plot illustrating time to gas ratio versus account cache hit
        rate.  This plot measures the effectiveness of maintaining an in-memory
        account cache. Lighter colored hexes indicate higher density of
        observation. The account cache is used by \texttt{BALANCE} and other
        opcodes that require loading an account into resident memory.  Notice
        that with a 1.0 cache hit rate, the time/gas ratio demonstrates greater
        variance than with 0 cache hits highlighting cache inefficacy.
    }}
\end{figure}

As a measure to smooth out this nondeterminism, \texttt{cpp-ethereum} implements
a memory-resident cache which provides temporary caching of recently accessed
state objects (\eg~accounts, code sizes, etc) as discussed earlier in
\Cref{sec:background}.  Not to be confused with the LRU Cache used by
LevelDB, the caching policy for these memory-resident caches
is determined by the Ethereum client;  whereas what
gets stored in the LevelDB cache is not controlled by the client. Due to frequent
state changes following each block, the custom implemented account cache is
invalidated when state is finalised.  To study the effectiveness of the
memory-resident cache implementation, we measured the account cache hit rate
which tracks the percentage of cache hits when the EVM runtime attempts to
lookup an account in the state trie.  It is used for all opcodes that
require an account lookup such as BALANCE.\@ Our measurements in
\Cref{fig:account-call-hitrate} show that the account cache is mildly effective
at improving the Time-to-Gas ratio.  On machine B, the advantages of having
the account cache appears marginal.

Cache efficiency is typically realised when programs have strong locality of
reference. We observed that during the runtime of smart contracts in our
workload, usage of the \texttt{BALANCE} opcode often referenced different
account balances which resulted in poor spatial locality. On the other hand, we
observe that the use of caching for \texttt{EXTCODESIZE} to be more effective:
the majority of the ratio distribution is below the baseline (1,000 ns/gas),
with
a mean that is below 100 ns/gas. We attribute this improvement to two usage
traits.  Firstly, the \texttt{EXTCODESIZE} opcode is often used to check code
sizes of well-known address locations.  Secondly, the code size for an address
is unlikely to change once code has been authored for that address:
Ethereum smart contracts are immutable until the
\texttt{SUICIDE} opcode is called (which deletes them).
In contrast, account balances change often, which causes cached values
in the account cache to be frequently invalidated reducing cache
effectiveness.
Therefore, we emphasize that memory-resident caching appears to be
sub-optimally effective to improve EVM execution performance for opcodes
that require account lookups.

\paragraph{Hardware Choices}
\begin{figure*}[htbp]
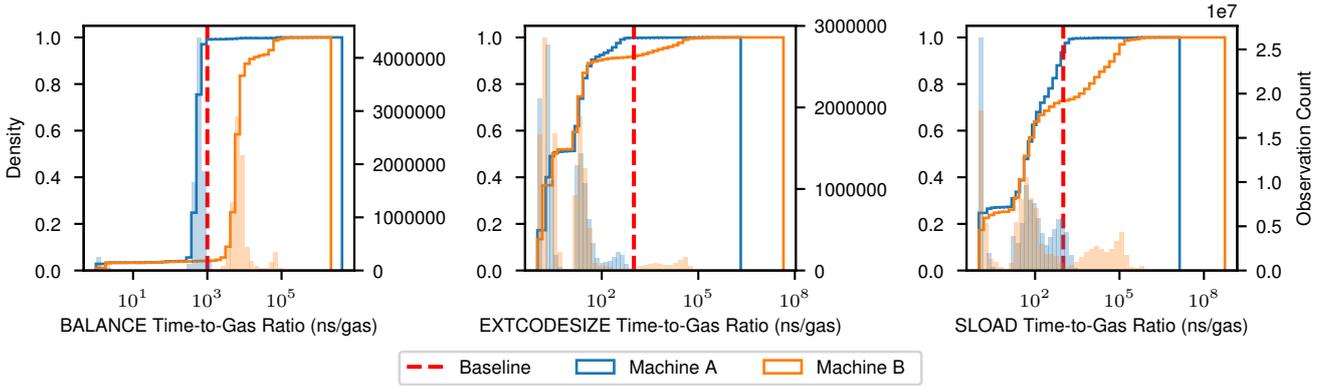

  \InputIfFileExists{plots/MISALIGNMENT.pgf}{}{\textbf{!! Missing Graphics !!}}
  \caption{Distribution of \texttt{BALANCE}, \texttt{EXTCODESIZE} and
  \texttt{SLOAD} time-to-gas ratios\label{fig:misalignment} showing that Machine
  B contains execution instances of the \texttt{BALANCE} opcode running longer
  than execution instances on Machine A by a factor of 10 to 100. Increasing
  time-to-gas ratio represents increasing unit time spent per unit gas.  The
  degree of misalignment is measured by the amount of ratios observed above (to
  the right of) the baseline. Hardware choice effects are measured by the union
  of overlapping distribution of ratio values.  Evidently, hardware choice plays
  a significant factor in the runtime of \texttt{BALANCE} and also causes a
  misalignment in its gas cost. Some minor misalignment as a consequence of
  hardware choice is also observed for both \texttt{EXTCODESIZE} and
  \texttt{SLOAD} for Machine B.}
\end{figure*}

\begin{figure}[htbp]
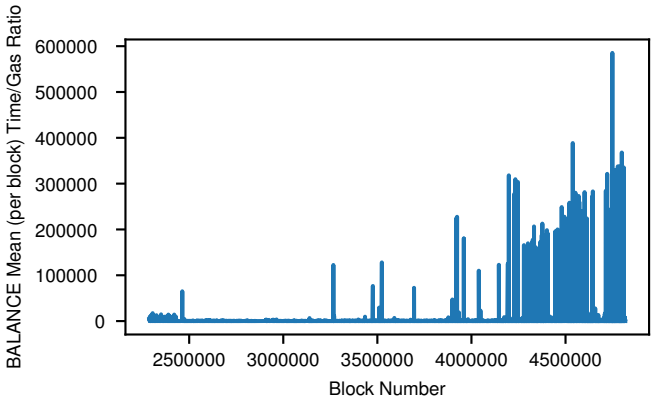

  \InputIfFileExists{plots/BALANCE-TIME.pgf}{}{\textbf{!! Missing Graphics !!}}
  \caption{Plot showing deterioration of \texttt{BALANCE} opcode as block number
  increases. As the state size grows, \texttt{external} opcodes efficiency
  deteriorate due to increased read amplification when accessing data values from
  disk.\label{fig:deterioration}}
\end{figure}
Choice of hardware when running an Ethereum node plays a significant factor for
performance and the efficacy of keeping up with consensus.  We observed that
despite running with competent consumer hardware machines (\ie Machine B), it
is still possible for the node to lag behind consensus as more capable machines
(likely to be used by miners) can outperform less capable machines.
Significant mean and $\sigma$ (mean absolute deviation)
differences for identical opcodes on the two machines
were observed.  Generally, both mean and mean absolute deviation observed on
Machine B are higher than observed on Machine A.  This deviation
highlights that machines with better hardware specifications are likely to
outperform lesser capable machines.
This is especially problematic as miners
are usually well-resourced, running nodes on powerful machines. Such miners
will have relatively better gas economy while executing EVM
transaction in contrast to \eg machines used in domestic households.

We argue that
this competitive advantage in gas economy for the well-resourced  poses a threat to node diversity
and decentralisation. Specifically,
it indicates that mining nodes with better hardware will gain a competitive edge
while extending the canonical chain, while nodes running on lesser hardware
will instead contribute to higher fork occurrences.
Nodes with similar specifications to Machine B
are less likely to keep up with consensus, especially if (as indicated in
\Cref{fig:deterioration}) gas economy deteriorates over time due to increased
read amplification.
In light of this, we suggest that consensus networks consider recommending
minimal hardware requirements for effective participation in consensus.

Significant deviation in performance due to hardware is particularly observable
for the \texttt{BALANCE} opcode, whose distribution of observed Time-to-Gas
ratios varies greatly between Machine A and Machine B (see
\Cref{fig:misalignment}).  The difference is so large that in the event that
both machines were operated by miners, we contend that
Machine B is almost guaranteed to
lose out in the race for consensus.
The difference between the two machines in terms of
their ratio distribution further confirms that hardware capability is a key
factor in the gas economy of a node.  Furthermore, we also observed that
Machine A and B have distinct runtime profiles.  Nearly all \texttt{BALANCE}
opcode executions run below the baseline ratio for Machine A, while only 10\% of
\texttt{BALANCE} executions run below the baseline for Machine B.

As a general observation, we observed that time-to-gas ratios on Machine B are
generally higher than Machine A and this difference is more profound for
\texttt{external} class opcodes.  Other opcodes show negligible difference and
still run below the baseline.  This difference further confirms that hardware
choice does play a significant role in an operator's efficiency and tendency to
keep up to consensus.

Aside from the mean ratio differences between the two machine, we also observe
that on average the assigned gas cost for non-\texttt{external} opcodes are
higher than \texttt{external} opcodes.  Therefore, it appears as if
users have been paying relatively more
in Gas for computational opcodes compared to \texttt{external} opcodes.

The observed variation in ratio distributions between the two machines
indicates that the Gas cost model is ineffective for maintaining
fairness between nodes running different hardware.
We argue that, due to this variation,  the Ethereum
ecosystem will tend to centralization as lesser hardware has relatively
fewer incentives to race for consensus.
The reduced incentives for lesser nodes to
participate increases the barriers to entry for consensus participation
and, we argue, so threatens node diversity. It also reduces the potential
for network
decentralization, since lesser capable nodes must either give up on verifying
future transaction executions or fall out of consensus.

%% file: content/solution.tex
\section{Possible Solutions}
In the previous section, we identified key problems that led to highly volatile
time-to-gas ratio distributions for various EVM opcodes.  These key problems are
mainly exacerbated by the dependency of transaction execution on state.  As a
direct measure to reduce this variance in ratio distribution, we can either
reduce state access during transaction runtime or improve I/O performance and
predictability when
state is being accessed.  In this section, we outline existing research and
potential implementation avenues that can be applied to address this problem.

One way improve I/O performance when reading state data is to optimize the
database backend used to store and read state data. Volatile database lookup
times suffer from read-amplification introduced by LevelDB storage.
Read-amplification occurs when the number of read operations executed by the
database driver exceeds a single query against the disk. \cite{conf/hotstorage/RajuPKOKCA18}
proposes a novel database implementation to store the Ethereum state objects
which improves read and write performance by reducing read and write
amplification.

Longer term solutions that aim to curb the state growth on a single node include
blockchain sharding and state channels. State channels
enable the aggregation of
transaction executions that persist all interim state changes in a single
transaction on the blockchain.  Proposed state channel solutions are similar to micro-payment
channel schemes proposed in Bitcoin such as
\cite{decker2015duplex,poon2015lightning} but envisioned to be more generic to
allow for more general purpose computation to take place between two private
parties.  Blockchain sharding solutions such as
\cite{al2017chainspace,kokoris2018omniledger,luu2016secure,sel2018towards} aim
to to split the network into shards (\ie subset of nodes) that are responsible
for a subset of the full state.  While not directly addressing the key problems
discussed in the previous section, proposed sharding scheme can alleviate the
problem by curbing node state growth.  Sharding schemes introduce additional security
assumptions that enable nodes to process state changes that they cannot
independently verify.

Finally, verifiable computing methods~\cite{ben2013snarks,bowe2018zexe} have also been proposed as a potential solution to avoid
the need for transaction execution. Such schemes allow a node to
verify that a transaction has been correctly executed without having to
re-execute the transaction to independently derive its result,
using embedded zero-knowledge proofs. These proofs can be checked in
time independent of the size of the transaction. Yet at the time of writing
have yet to be deployed.

%% file: content/related.tex
\section{Related Works}
\label{sec:related}

Our work is closely related to~\cite{8612882},
which briefly measured the performance of a handful of Ethereum opcodes,
and compared the results between two similar machines running different
operating systems. None of the opcodes considered in their analysis
access the state trie (\ie perform I/O), and their measurements cover
synthetic benchmarks only for a relatively small number of executions.
Thus, unlike our results, theirs do not easily
generalise to real-world usage nor shed light on the impact of I/O on
the current Gas cost model, which we found to be a major source of
misalignment. Finally, their results shed little light on the plight of
modest nodes vs their more computationally resourced peers, which we identify
as a threat to node diversity.

Empirical measurements have provided us with a better understanding of
decentralized cryptocurrency networks.  \cite{decker2013propagation} measured
block and transaction propagation delays in Bitcoin and produced a model which
led to future security studies and the discovery of numerous attacks such
as~\cite{eyal2018majority,nayak2016stubborn,sapirshtein2016optimal}.
~\cite{2018arXiv180103998E} measured decentralized metrics in both Bitcoin and
Ethereum networks such as the distribution of hash power held by miners and the
distribution of network peers by geography, to quantitatively understand the
degree of decentralization.
\cite{2016arXiv160606530A,Kim:2018:MEN:3278532.3278542} extend our understanding
of the Ethereum network by adding measurements of client heterogeneity,
connectivity between peers and geographical distribution of peers.  Other
measurement and tracing work such as
\cite{fleder2015bitcoin,koshy2014analysis,ron2013quantitative} involve the
analysis of the security of the underlying network which led to a better
understanding of the decentralized network security model.  \cite{etherscan} is
a community-run Ethereum block explorer which also collects EVM traces for each
transaction. Our work extends these traces by also including measured time for
each EVM operation, and carefully analysing the effectiveness of the current Gas
cost model.

Other works have focused on smart contract use, \cite{bartoletti2017empirical}
quantified the number of smart contracts in application domains for usage
pattern analysis in Bitcoin and Ethereum.
\cite{Kiffer:2018:AEC:3278532.3278575} extend this by analyzing behavioral usage
patterns by network users and clustering contracts based on code similarity.

Efforts to improve verification throughput have prompted a flurry of studies in
speculative concurrent execution of EVM transactions.
\cite{dickerson2017adding,yu2017smart}~proposed adding concurrency in smart
contracts and showed that it was possible to execute smart contracts
concurrently by analysing transaction dependencies to run non-conflicting
transactions in parallel.  Subsequently, \cite{2018arXiv180901326S}~introduced
the first framework for optimistic concurrent execution of smart contracts which
leverages basic/multi-version timestamp ordering. Finally,
\cite{2019arXiv190101376S} carried out a measurement study on simulated greedy
concurrent transaction execution. Our measurements and analysis assume serial
transaction execution, as implemented in the main Ethereum clients; however
extending them to concurrent execution could be an interesting avenue for future
research.

%% file: content/conclusion.tex
\section{Conclusion}
\label{sec:conclusion}
To the best of our knowledge, we performed the first large scale empirical
analysis of the EVM Gas mechanism and the current Gas cost model.  We confirmed
potential denial-of-service vectors resulting from mis-priced EVM opcodes
involving disk I/O. We also measured the effects of I/O in transaction costs,
finding that it remains poorly captured by the current Gas cost model and that
its effects are getting worse over time.  We also found that under the current
Gas cost model, nodes with modest computational resources are disadvantaged
compared to their better resourced peers, which we identified as an ongoing
threat to node diversity and network decentralization. Our results indicate
that, absent deployment of techniques such as verifiable computation that allow
transactions to be verified without re-executing them, further work is required to improve Ethereum's Gas cost
model to better align Gas costs with transaction execution costs.